\documentclass[onecolumn,floatfix,superscriptaddress,a4paper,showpacs,showkeys,nofootinbib,notitlepage]{revtex4-1}
\usepackage[colorlinks=true,linktocpage=true,linkcolor=blue,citecolor=blue,allcolors=blue]{hyperref}
\usepackage{epsfig}
\usepackage{latexsym}
\usepackage[utf8]{inputenc}
\usepackage{xspace}
\usepackage{indentfirst}
\usepackage{enumitem}
\usepackage{color}
\usepackage{placeins}

\usepackage{setspace}
\usepackage{lipsum}

\usepackage{hyperref}

\usepackage{todonotes}

\usepackage{amsmath}
\usepackage{amssymb}
\usepackage[english]{babel}
\usepackage{url}
\graphicspath{{figs/}}
\newcommand{\mean}[1]{\langle #1 \rangle}
\newcommand{\eq}[1]{\begin{align} #1 \end{align}}

\newcommand{\be}{\begin{equation}}
\newcommand{\ee}{\end{equation}}

\begin{document}
\title{
Net-proton fluctuations influenced by baryon stopping and quark deconfinement
}
\author{Oleh Savchuk}\thanks{Corresponding author}
\email{savchuk@frib.msu.edu} 
\affiliation{Facility for Rare Isotope Beams, Michigan State University, East Lansing, MI 48824 USA}
\affiliation{Bogolyubov Institute for Theoretical Physics, 03680 Kyiv, Ukraine}

\date{\today}
\begin{abstract}

Preliminary data from the Beam-Energy Scan II~\cite{pandav} measurements by the STAR Collaboration at the Relativistic Heavy Ion Collider suggest a dip in the fourth-to-second-order cumulant ratio when plotted vs. beam energy. At the same energy range where the structure appears, a transition from hadrons to quarks is expected, the deconfinement transition. In this paper, the role of quark deconfinement in establishing fluctuaitions in the early stages of the collision is considered. Two models are compared: one with stopping occurring on a baryon-by-baryon basis, and a second where stopping proceeds through quark degrees of freedom. In the latter model, the fluctuation of baryon number is significantly reduced and this signal is found to survive recombination into hadrons and the subsequent diffusion. The transformation from baryon to quark stopping thus produces a dip in the fourth-to-second-order cumulant ratio when plotted vs. beam energy, consistent with observations.

\end{abstract}

\maketitle
\section{Introduction}
Fluctuations of conserved charges play a crucial role in the experimental search for the critical point of QCD using heavy-ion collisions~\cite{Stephanov:1998dy,Stephanov:1999zu,Athanasiou:2010kw,Stephanov:2008qz,PhysRevC.86.024904,Vovchenko:2015uda}. These fluctuations can act as proxies for phase transitions and interactions, thereby enhancing our understanding of the equation of state. They are sensitive to the initial stages of the collision, providing insights into charge stopping, momentum fluctuations, jets, and other phenomena~\cite{Stephanov:2009mz,Asakawa:2000wh}.

Recently, the STAR Collaboration released preliminary data on net-proton cumulants that show deviations from the global conservation baseline~\cite{Vovchenko:2021kxx,Braun-Munzinger:2020jbk}. This baseline assumes no correlation between protons in momentum and/or coordinate space. Any deviation from this baseline indicates an interaction~\cite{Vovchenko:2020tsr,Poberezhnyuk:2020ayn}. Non-trivial changes in fluctuations with energy could potentially signal the presence of the critical point\cite{Stephanov:1998dy,Stephanov:1999zu,Shuryak:2024yoq}.

Multiplicity fluctuations are typically characterized by central moments, such as \(\langle(\Delta N)^{2}\rangle \equiv \sigma^2\), \(\langle (\Delta N)^{3}\rangle\), \(\langle (\Delta N)^{4}\rangle\), etc., where \(\langle...\rangle\) denotes event-by-event averaging and \(\Delta N \equiv N -\langle N \rangle\). Intensive measures like the scaled variance \(\omega\), (normalized) skewness \(S\sigma\), and kurtosis \(\kappa\sigma^{2}\) are defined in terms of central moments as follows:

\begin{equation}
\omega[N] = \frac{\sigma^2}{\langle N \rangle} = \frac{\kappa_2}{\kappa_1},
\end{equation}

\begin{equation}
S\sigma[N] = \frac{\langle (\Delta N)^{3} \rangle}{\sigma^{2}} = \frac{\kappa_3}{\kappa_2},
\end{equation}

\begin{equation}
\kappa\sigma^{2}[N] = \frac{\langle (\Delta N)^{4} \rangle - 3 \langle (\Delta N)^{2} \rangle^2}{\sigma^{2}} = \frac{\kappa_4}{\kappa_2},
\end{equation}
where \(\kappa_n\) are the cumulants of the \(N\)-distribution.In this context, $N$ refers to the net-proton number, which is the total number of protons minus anti-protons within the kinematic acceptance of the detector for a given event class. Measuring baryon charge is challenging due to the fact that neutrons are not detectable by the detectors. However, the net-proton number serves as a good proxy for net-baryon fluctuations. This is because the relation between baryons and protons is governed by a binomial probability, considering isospin randomization~\cite{PhysRevC.86.024904}.

Recent predictions suggest that the critical point should be located in dense nuclear matter at lower collision energies (\(\sqrt{s_{NN}}<7.7~\text{GeV}\))~\cite{Sorensen:2024mry,Sorensen:2023zkk,Basar:2023nkp}. However, the discrepancy between the STAR data and the baseline occurs in the region of \(\sqrt{s_{NN}}=7.7-20~\text{GeV}\). Thus, this discrepancy might arise from phenomena not associated with the critical point~\cite{Holzmann:2024wyd,Poberezhnyuk:2019pxs}. In this energy range, degrees of freedom are expected to transition from hadrons to quarks, known as the onset of deconfinement. Signatures of this transition were found in studies by the NA61 collaboration~\cite{Gazdzicki2004, BraunMunzinger2007, Gazdzicki2011, Abgrall2014}. Another concurrent phenomenon in this region is string melting or recombination, as discussed in~\cite{Lin2005, XuGreiner2005, Lin2014,Fries:2003vb,Fries:2003kq,Fries:2004ej,Muller:2022htn}. 

Since the number of degrees of freedom differs by a factor of three between quarks and baryons, we expect significant differences in the fluctuations of conserved charges. Specifically, if one considers random stopping of quarks instead of random stopping of baryons the scaled variance and skewness are reduced by a factor of three, and the kurtosis by a factor of nine. At higher beam energies, fluctuations are dominated by baryon-antibaryon pair creation. In this regime, fluctuations are understood by considering the effects of local charge conservation~\cite{PhysRevC.102.044909,PhysRevC.102.064906,Savchuk:2023yeh,Savchuk:2024ycj,Braun-Munzinger:2023gsd} projected onto the acceptance.

In this paper, I demonstrate that transitioning from a scenario where baryons are stopped independently to one where quarks are stopped independently and then recombine into baryons leads to a pronounced dip in the baryon kurtosis at the transition energy. By incorporating a model where recombined baryons either enter or exit the acceptance, along with a simple parameterization of the contribution from pair creation, I find that the dip in kurtosis persists but is reduced in magnitude. Meanwhile, the skewness and scaled variance become nearly featureless. The resulting scaled moments are quantitatively consistent with STAR results.

\section{Baryon stopping}\label{sec 1}

Baryon charge and its fluctuations can be created in heavy-ion collisions through two distinct processes: as the stopped charge carried by colliding nuclei, and as particle-antiparticle pairs created during the collision. The stopped charge can occupy the region at midrapidity, where the measurements are performed, and influence fluctuations. At lower energies, this is the dominant effect in baryon charge fluctuations. In the following, we will assume an independent and random distribution of this charge in space.
Let the function \( P(B,\bar{B}) \) be a normalized probability distribution for emitting \( B \) baryons and \(\bar{B} \) anti-baryons in the "full" phase space. I will reconstruct this probability as follows:
\begin{equation}\label{full-space}
P(B,\bar{B}) = \delta(B - B_{s} - \bar{B}) P(\bar{B}, B_{s}),
\end{equation}
where the number of baryons \(B\) consists of baryons stopped \(B_s\) in a collision and the number of baryons created in pairs with anti-baryons \(\bar{B}\). In the most central collisions, one might assume that:
\begin{equation}\label{full-stop}
P(\bar{B},B_s) = \delta(B_s - 2A) P(\bar{B}),
\end{equation}
i.e., the number of stopped baryons is fixed to the total number of baryons in the colliding nuclei \(2A\) and does not fluctuate.

To connect the fluctuations in different rapidity intervals, I assume that the acceptance of particles is binomial. This means that each particle of a given type is accepted by the detector with a fixed probability \(\alpha\) \cite{PhysRevC.86.044904,PhysRevC.85.021901}. This probability, \(0 \leq \alpha = \langle n \rangle / \langle N \rangle \leq 1\), equals the ratio of the mean number \(\langle n \rangle\) of particles accepted in a fixed region of momentum space \(\Delta y\) to the mean number \(\langle N \rangle\) of particles of the same type in a "full" momentum space \(\Delta Y\). Full momentum space does not necessarily mean complete ``\(4\pi\)" acceptance. The sufficient condition for \(\Delta Y\) is to fully encompass \(\Delta y\)~\cite{Savchuk:2022ljy}. The main assumption of binomial acceptance is that the probability \(\alpha\) is the same for all particles of a given type and is independent of any properties of a specific event. This assumption allows us to relate the cumulants within a finite acceptance to their values in the larger, encompassing phase space.

The binomial procedure is also sensitive to the definition of the particle. Thus, one expects different results depending on whether I apply it to hadrons or quarks. In the case of hadron charge carriers, the conditional probability for observing \( b \) baryons and \(\bar{b} \) anti-baryons using Eqs. (\ref{full-space}) and (\ref{full-stop}) within a finite binomial acceptance is given by:
\begin{equation}\label{dist}
p(b,\bar{b}|\alpha) =  \delta(b-b_{s}-b_p)P(b_s|\alpha_s)P(b_p,\bar{b}|\alpha_p)=\delta(b-b_{s}-b_p)B(B_s,b_s|\alpha_s)P(b_p,\bar{b}|\alpha_p),
\end{equation}
where I have introduced \(\alpha_s\) and \(\alpha_p\) as the acceptance probabilities for particles that were originally stopped and those created in pair production, respectively. \( B(B_s, b_s | \alpha_s) \) represents the binomial distribution for baryons that were stopped inside the acceptance \( b_s \), given that the total stopped charge is \( B_s \):
\begin{equation}\label{binomial}
B(N,n|\alpha)=   
 \frac{N!}{n!(N-n)!}\,\alpha^n (1-\alpha)^{N-n}.
\end{equation}
One can therefore see from Eq. (\ref{dist}) that the distribution of stopped baryons and charge pairs can be studied separately, \(\kappa_n=\kappa_n^{\rm stopping}+\kappa_n^{\rm pair}\). In a later section, I will consider the contributions from pair creation. Cumulants of stopped hadrons are:
\begin{equation}\label{stop-mean}
\kappa_1^{\rm stopping} = \alpha_s B_s,
\end{equation}
\begin{equation}\label{stop-var}
\kappa_2^{\rm stopping} = B_s(1-\alpha_s)\alpha_s,
\end{equation}
\begin{equation}\label{stop-skew}
\kappa_3^{\rm stopping} = B_s\alpha_s(1-\alpha_s)(1-2\alpha_s),
\end{equation}
\begin{equation}\label{stop-kurt}
\kappa_4^{\rm stopping} = B_s\alpha_s(1-\alpha_s)(1-6\alpha_s(1-\alpha_s)).
\end{equation}

\section{Quark Stopping}\label{sec 2}

On the other hand, it is possible that stopped charge is carried by quarks that later recombine into hadrons~\cite{Fries:2003vb,Fries:2003kq,Fries:2004ej,Muller:2022htn}. In this case an emitted baryon might be comprised by quarks taken from several different incoming baryons. First, one can consider a number of stopped baryon number $b_s$, i.e. an amount of baryon number in some designated mid-rapidity region of coordinate space. Because the stopped baryon number involves three times as many independent choices compared to the baryon stopping picture of the previous section, the cumulants for the stopped number of baryons, $\kappa_n$, are smaller by a factor of $3^{-(n-1)}$. However, once the quarks recombine, some baryons from the designated mid-rapidity region might be emitted outside the rapidity window, while others from outside the region might be emitted into the acceptance window. This effect, which can also be enhanced by diffusion, will tend to shift the cumulants toward those of the baryon-stopping picture. Thus, we consider a more involved model for this section. A simple binomial model, utilizing a single parameter $\alpha_s$, is applied to the stopping describing the number of quarks in the designated region:
\eq{\label{quark-st}
P(b_s) = B(3B_s, 3b_s | \alpha_s),
}
and a transfer matrix with two independent parameters is then applied to describe the subsequent movement of baryons.

\subsection{Recombination}
\begin{figure}
\includegraphics[width=.49\textwidth]{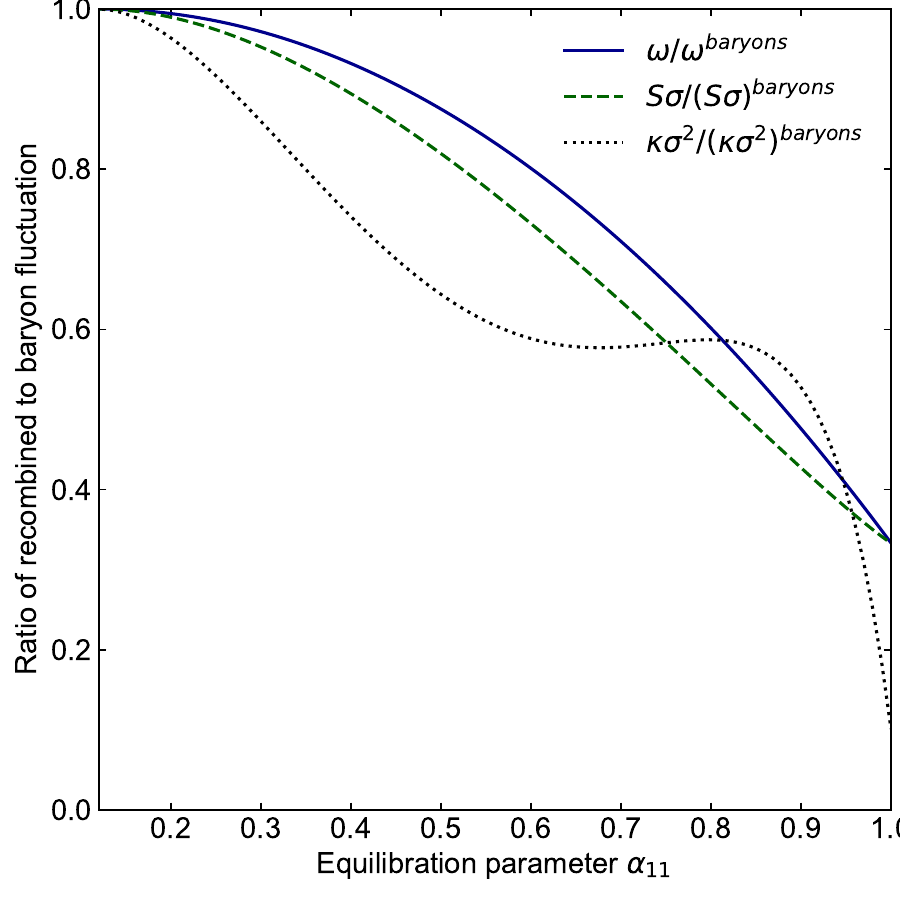}
\caption{\label{recombination}Scaled variance, skewness and kurtosis of stopped hadrons divided by values for stopped quarks. The values of fluctuation measures move from quarks to hadronic baseline as equilibration parameter decreases. Certain scalling on the order of cumulant is being observed.}
\end{figure}

The rapidity space can be split into rapidity intervals \(\Delta y_i\) large enough so that recombination happens in each of these intervals in a way that charge is conserved, i.e., there is no need to involve quarks from neighboring cells. Supposedly, after recombination, each hadron stochastically moves independently of all others, leading to the binomial probability of moving from the rapidity interval \(i\) to \(j\): \(\alpha_{i j}\) over some time. Then the distribution of particles at the point \(i\) becomes:
\eq{
P^t(b_s^{i})=\delta\left(\sum_j b_s^{ji} - b_s^i\right)B(b_s^{ji}, b^{0 j}_s | \alpha_{ji}) P^0(b^{0 j}_s),
}
where \(b_s^{ij}\) stands for the number of hadrons that moved to \(j\) from \(i\), \(\alpha_{ij}\) represents probabilities of such transitions, and \(P^0(b^{0\,i}_s)\) is the joint probability of stopped charges right at the transition from quarks to hadrons.
The generating function of the final distribution then becomes:
\eq{
F_i^t = \langle \exp(k b_s^i) \rangle = F^0(\phi(k| \alpha^t_{ji})),
}
where \(F^0\) is the cumulant generating function of \(P^0(b^{0\,i}_s)\) and \(\phi[k|\alpha] \equiv \ln(1 - \alpha + \alpha e^{k})\).

Cumulants of the joint distribution can be defined as follows:
\eq{
\kappa_{i_1 \cdots i_n} = \frac{1}{n!}\frac{\partial^n F\left(\vec{k}\right)}{\partial_{i_1} \cdots \partial_{i_n}},
}
where the indices \(i\) correspond to any of the parameters in \(\vec{k}\). Using this definition, one arrives at:
\eq{\label{rec-mean}
\kappa_i &= \kappa^0_j \alpha_{j i},\\ \kappa_{ii} &= \kappa^0_j \alpha_{j i} (1 - \alpha_{j i}) + \kappa^0_{jk} \alpha_{j i} \alpha_{k i},\\
\kappa_{iii} &= \kappa^0_j \alpha_{j i} (1 - \alpha_{j i}) (1 - 2\alpha_{j i}) 
+ 3 \kappa^0_{jk} \alpha_{j i} (1 - \alpha_{j i}) \alpha_{k i} 
+ \kappa^0_{jkl} \alpha_{j i} \alpha_{k i} \alpha_{l i},\\
\label{rec-kurt}\kappa_{iiii} &= \kappa^0_j \alpha_{j i} (1 - \alpha_{j i}) (1 - 6\alpha_{j i} (1 - \alpha_{j i})) \\ &+  \kappa^0_{j k} \left[ 4\alpha_{k i}\alpha_{j i} (1 - \alpha_{j i}) (1 - 2\alpha_{j i}) + 3\alpha_{k i} \alpha_{j i} (1 - \alpha_{j i}) (1 - \alpha_{k i}) \right] \nonumber \\ &+ 2\kappa^0_{jkl} \alpha_{j i} \alpha_{k i} \alpha_{l i} \left[(1 - \alpha_{j i}) + (1 - \alpha_{k i}) + (1 - \alpha_{l i})\right] \nonumber \\ 
&+ \kappa^0_{jklm} \alpha_{j i} \alpha_{k i} \alpha_{l i} \alpha_{m i},\nonumber
}
where \(\kappa^0\) and \(\kappa\) are cumulants of quarks and hadrons, respectively.

For simplicity, I will consider two subsystems that correspond to rapidity windows inside ``1" and outside ``2" of acceptance. This requires us to define \(a_{11}, a_{22}\) as \(a_{12} = 1 - a_{11}\) and \(a_{21} = 1 - a_{22}\). The number of parameters can be decreased to one if I also assume that the mean charge remains the same:
\[
\alpha_{21} = \frac{\kappa_1}{\kappa_2}(1 - \alpha_{11}) = \frac{\kappa_1}{\kappa_2}a_{12},
\]
which reminds us of the detailed balance condition. This should be satisfied if rapidity spectra remain constant. The initial \(\kappa^0_1\) and \(\kappa_2^0\) can be different; however, they will quickly reach equilibrated values. For simplicity, I will assume \(\kappa^0_i = \kappa_i\). This allows us to study the dependence of cumulants on \(a_{11}\) alone. In equilibrium, this \(a_{11}\) should be proportional to the subvolume-to-volume ratio, which can be estimated as follows:
\[
a^{eq}_{11} = \frac{\kappa_1}{\kappa_1 + \kappa_2}.
\]
If the particle lacks time to diffuse, \(a_{11} = 1\). Thus, one expects that over time \(a_{11}\) will decrease in the interval \(\frac{\kappa_1}{\kappa_1 + \kappa_2} \leq a_{11} \leq 1\).

In what follows, I will assume that the number of stopped quarks is binomially distributed between \(1\) and \(2\) with \(\alpha\) that corresponds to the one extracted from STAR data, building $P^0(b_s^{0\,j})$ from Eq.(\ref{quark-st}). After that, Eqs. (\ref{rec-mean}-\ref{rec-kurt}) are applied in order to obtain the cumulant dependence on \(\alpha_{11}\). In order to simulate the proton observable, \(\alpha_{ij}\) can additionally be multiplied by the probability of a baryon being a proton.

Figure (\ref{recombination}) shows the dependence of scaled variance, skewness, and kurtosis ratio of hadrons and quarks as a function of \(a_{11}\) using \(\alpha \approx 0.1\), \(B_s = 337\) in Eq.(\ref{quark-st}) for initial quark redistributions $P^0(b_s^{0\,j})$ and Eqs.(\ref{rec-mean}-\ref{rec-kurt}). The acceptance effects of quarks and baryons cancel in the ratio and the limiting values correspond to grand canonical ensemble values, $\alpha_{11}=1$ or $\alpha_{eq}$. One can see that when \(\alpha_{11} = 1\), simultaneously \(\alpha_{22} = 1\), and fluctuations of quarks remain unchanged after recombination into hadrons. However, when \(\alpha_{11} = \alpha_{11}^{eq}\), hadronic values are recovered and the signal of the quark phase is lost. This recovery happens faster for scaled variance and slower for skewness. Kurtosis at small values of \(a_{11}\) shows the slowest approach to hadronic values. However, a certain non-monotonic structure is present at \(\alpha_{11} \approx 0.8\). It is interesting to consider the implications of this peculiar behavior of kurtosis.

This behavior of cumulants suggests that it is possible to obtain values of cumulants that correspond to a non-interacting baseline in lower orders but preserve effects of interactions in higher ones. It has also been observed in transport studies that scaled variance and skewness depend weakly on the potential interaction between baryons, while kurtosis shows notable differences~\cite{PhysRevC.107.024913}. Recent results of the Beam Energy Scan II at STAR, where scaled variance and skewness are well described by a non-interacting baseline (which here corresponds to baryon stopping with pair creation) but not kurtosis, can be understood with this simple model of independent charge movement. It is also interesting to apply this approach as an aproximation
of the hadronic stage following ``freeze-out" fluctuations~\cite{Pratt:2017oyf,Pradeep:2022mkf}.

\section{Pairs}\label{sec 3}
Both stopping scenarios should have identical pair production mechanics. The pair contribution is mostly featureless~\cite{PhysRevC.102.064906}, and in the energy range where the structure of interest appears, it is much weaker compared to the stopped charge contribution. It can also be affected by local conservation effects~\cite{Braun-Munzinger:2023gsd} seen in the balance function calculations~\cite{PhysRevC.102.044909,PhysRevC.102.064906,Savchuk:2023yeh,Savchuk:2024ycj}. In what follows, it is assumed that pairs are produced independently from one another and from the stopped charge.
 Pair multiplicity in the whole event can be correlated with the amount of stopped charge. Due to baryons and anti-baryons being created at the same rapidity \(y\), their final probability of being accepted by the detector will depend on the point where the pair is created, denoted as \(a_p(y)\). Therefore, it is beneficial to introduce \(P(\bar{B},y)\), the probability of creating a baryon and anti-baryon pair at point \(y\), with the total number of pairs being \(\bar{B}\). The relationship between fluctuations in \(y\) and the "full" phase space is given by $P(\bar{B}) = \int dy\,P(\bar{B},y).$ The acceptance parameter for pairs created later in the collision depends on \(y\), denoted as \(\alpha=\alpha(y)\). Pairs initially created within acceptance tend to remain there, reducing the production of charge fluctuations:

\begin{equation}\label{dist-stop}
P(b_p,\bar{b}|\alpha_p) = \sum_{\bar{B}=0}^{\infty} \int dy\, P(\bar{B},y) B(\bar{B},b_p|\alpha_p(y)) B(\bar{B},\bar{b}|\alpha_p(y)).
\end{equation}
Assuming \(\alpha_p(y)=\alpha_p\) cumulants from pairs can be written as follows:
\begin{equation}\label{pair-mean}
\kappa_1^{\rm pair} = 0,
\end{equation}
\begin{equation}\label{pair-var}
\kappa_2^{\rm pair} = 2\mean{\bar{B}}(1-\alpha_p)\alpha_p ,
\end{equation}
\begin{equation}\label{pair-skew}
\kappa_3^{\rm pair} = 0,
\end{equation}
\begin{equation}\label{pair-kurt}
\kappa_4^{\rm pair} = 2\mean{\bar{B}}\alpha_p(1-\alpha_p)(1+6\alpha_p(1-\alpha_p)(\omega[\bar{B}]-1)).
\end{equation}
In the case of equal acceptance parameters, \(\alpha_s = \alpha_{p} \equiv \alpha\) , cumulants of stopped baryons and pairs reduce to the known result~\cite{PhysRevC.101.024917,PhysRevC.86.024904,PhysRevC.87.014901}:
\begin{equation}\label{global-var}
\omega = \frac{\langle B + \bar{B} \rangle}{\langle B - \bar{B} \rangle} (1 - \alpha),
\end{equation}
\begin{equation}\label{global-skew}
S\sigma = \frac{\langle B - \bar{B} \rangle}{\langle B + \bar{B} \rangle} (1 - 2\alpha),
\end{equation}
\begin{equation}\label{global-kurt}
\kappa\sigma^2 = 1 + 3\alpha(1 - \alpha) \left( \omega[\bar{B}+B] - 2 \right).
\end{equation}
In case of quark stopping one has to add cumulants after recombinations Eqs.(\ref{rec-mean}-\ref{rec-kurt}) to cumulants of pairs Eqs.(\ref{pair-mean}-\ref{pair-kurt}).

\section{ STAR results}\label{STAR}

\begin{figure}
\includegraphics[width=.49\textwidth]{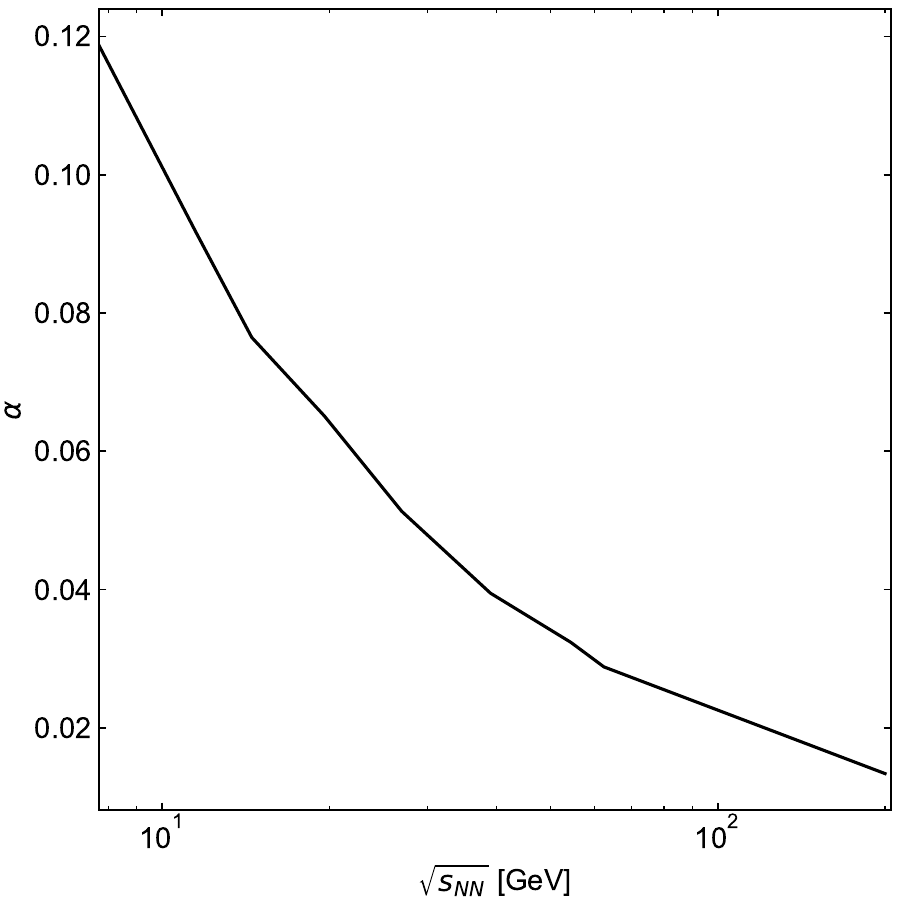}
\caption{\label{alpha} The acceptance parameter is plotted versus collision energy. The acceptance correction here includes both kinematic acceptance and species selection. Kinematic acceptance is caused by $p_T$ and $y$ binning of the data. The probability of baryon being a proton is set to $40\%$.}
\end{figure}

To describe the proton number fluctuations in Au+Au central collisions at \(\sqrt{s_{\rm NN}} = 7.7-200~{\rm GeV}\) as measured by the STAR Collaboration~\cite{pandav}, I use the baryon or quark stopping and pair production outlined in Secs.~(\ref{sec 1}-\ref{sec 3}). The STAR data for \(\omega\), \(S\sigma\), and \(\kappa \sigma^2\) were measured at 5\% central collisions and are presented for a symmetric rapidity interval \(\Delta y = 1\) in the center of mass system. A transverse momentum cut \(0.4 < p_T < 2.0~\text{GeV}/c\) was applied.

The first step of the comparison is determination of binomial probability  \(\alpha\) for quark or baryon to appear inside acceptance for different collision energies. Since the number of anti-baryons equals the number of pairs, one expects that net charge would be a measure of stopped charge. Thus, the acceptance \(\alpha_s\) for stopped protons, assuming binomial probability of baryon being proton, should be equal to:
\begin{equation}\label{as}
\alpha_s = \frac{\langle p - \bar{p} \rangle_{|y| < 0.5}}{\langle B - \bar{B} \rangle},
\end{equation}
where \(\langle B - \bar{B} \rangle\) in the full space is equal to the number of participating nucleons. The values of these parameters were reported by the STAR Collaboration alongside net-proton multiplicity~\cite{STAR:2020tga,STAR:2021iop}.

On the other hand, \(\alpha_p\) is much harder to estimate. Knowledge of the number of anti-baryons in the full ``$4\pi$" acceptance is required. Usually, this can be done using anti-proton spectra from hydrodynamics or fits of the data. In principle, factorial cumulants of protons and anti-protons can be used for a robust determination of \(\alpha_s\) and \(\alpha_p\). Generally, this leads to \(\alpha_p = (1-5)\alpha_s\) depending on the collision energy. It is also possible that the pair creation mechanism is influenced by local conservation, resulting in the volume of the system where charge is conserved being smaller than the system itself. This would translate into \(\alpha_p\) being larger compared to the one obtained from spectra, with \(\alpha_p(y)\) being sensitive to the rapidity \(y\) from which the pair originated. In my analysis, both scenarios include identical pair contributions; thus, the effect of hadronic vs. quark stopping should be independent of the pair contribution. The simplest estimation used seems to be robust when used to estimate charge fluctuations.

Once \(\alpha=\alpha_s=\alpha_p\) is determined, the mean numbers of baryons and anti-baryons can be calculated as:
\begin{equation}\label{b}
\langle B \rangle = \frac{\langle p \rangle_{|y| < 0.5}}{\alpha},
\end{equation}
\begin{equation}\label{bbar}
\langle \bar{B} \rangle = \frac{\langle \bar{p} \rangle_{|y| < 0.5}}{\alpha}.
\end{equation}
This information is sufficient to calculate the scaled variance and skewness of the data. To calculate kurtosis, \(\omega[\bar{B} + B]\) is also required. I used the ideal gas model that reproduces \(\langle B \rangle\) and \(\langle \bar{B} \rangle\) from Eqs.~(\ref{b}) and~(\ref{bbar}) in the canonical ensemble to determine \(\omega[\bar{B} + B]\). However, the values of \(\omega[\bar{B}+B]\) obtained from the ideal hadron gas model lead to around a 5\% variation in the observable compared to \(\omega[\bar{B} + B] = 1\). Thus, this effect can be safely ignored.
\begin{figure}[h!]
\includegraphics[width=.49\textwidth]
{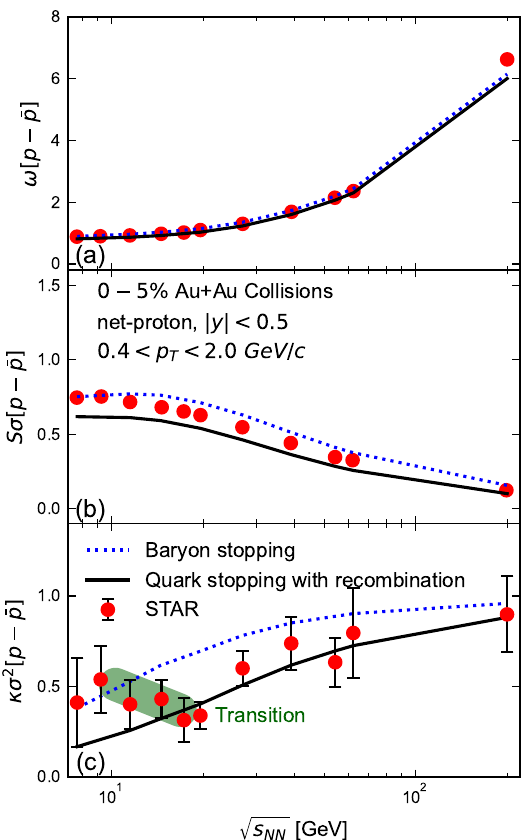}

\caption{\label{fluks} Scaled variance (a), skewness (b), and kurtosis (c) of the proton number distribution are plotted as functions of collision energy from preliminary STAR data. The dotted blue line corresponds to my simplified calculation of the canonical baseline. The solid black line shows the effect of quark stopping on fluctuation measures. Each line includes contributions from pair creation. The green region shows a possible change in kurtosis as degrees of freedom transform from hadrons to quarks and deconfinement occurs. Expected quark fluctuations are smaller than baryon ones. However, due to recombination, they do not survive in scaled variance (a) and skewness (b).
}
\end{figure}

The resulting values of fluctuation measures as functions of collision energies are shown in Figure~\ref{fluks}. Two models were used to describe the data. The first model assumes that the stopped charge is described as baryons with a binomial probability to occupy mid-rapidity, as described in Section \ref{sec 1}, Eqs.~(\ref{stop-mean}-\ref{stop-kurt}). The second model describes the stopped charge as being carried by quarks with a binomial probability to occupy mid-rapidity. However, in this case, a recombination model is applied that accounts for the random movements of quarks recombined into baryons, as described in Section \ref{sec 2}. To calculate hadronic cumulants after recombination, \(\alpha_{11} = 0.75\) is used, as in Eqs.~(\ref{rec-mean}-\ref{rec-kurt}), independent of collision energy. In every case, an additional contribution is included that corresponds to pair creation, Section \ref{sec 3}, Eqs.~(\ref{pair-mean}-\ref{pair-kurt}). Baryons/quarks and anti-baryons then enter acceptance with \(\alpha\) extracted from Eq.~(\ref{as}). The energy dependence of \(\alpha\) is shown in Fig.~\ref{alpha}.

The pair production contribution increases with collision energy, leading to both models providing similar expectations at \(\sqrt{s_{NN}}=200~\text{GeV}\). Some investigation was done into the \(\alpha_p(y)\) and deviations from \(\alpha_s\). Since acceptance parameters are already small, especially at the energies when pair production dominates, the variation of \(\alpha_p\) does not lead to substantial modification of kurtosis. However, this might become important once rapidity window \(\Delta y\) dependencies are considered, especially for factorial cumulants of protons and anti-protons. This is outside the scope of the current paper.

As expected, quarks have smaller values of fluctuations compared to hadrons. When ratios are taken, conservation laws factor out, and the scaled variance and skewness decrease by a factor of three, while kurtosis decreases by a factor of nine. Recombination leads to low-order cumulants approaching hadron values, as seen in scaled variance and skewness, mitigating this suppression. The situation is much better for kurtosis, which still retains most of the quark fluctuations.

The scaled variance shows a monotonic increase with collision energy. This is well described by the models. Baryon stopping slightly overestimates the data. On the other hand, quark stopping with recombination  preserves some of the suppression of fluctuations expected from quarks and lies closer to the data points. Both models fail to describe the \(\sqrt{s_{NN}} = 200~\text{GeV}\) point, which might indicate the oversimplified model for pair correlation, such as annihilation~\cite{Savchuk:2021aog,Pratt:2022kvz,Vovchenko:2022xil}.

Since odd cumulants of the pair production part of the distribution vanish in Eq.~(\ref{pair-skew}), skewness appears to be the most sensitive observable with regard to the stopped charge over all energy ranges. The hadronic stopping scenario overestimates the data, while quark values with recombination provide smaller skewness values. Both scenarios provide a description of similar quality except for the lowest energies \(\sqrt{s_{NN}}<14~\text{GeV}\), where the hadronic scenario performs better.

So far, both models provided similar behavior in comparison to the measured fluctuations. Thus, kurtosis should be used to distinguish between them. An interesting trend is observed in kurtosis, which shows a dip when compared to the global conservation baseline, corresponding to the baryon stopping scenario. Starting from the dip, quark stopping starts to describe the data. Notably, baryon stopping performs better at collision energies \(\sqrt{s_{NN}} < 14~\text{GeV}\), and quark stopping at \(\sqrt{s_{NN}} > 14~\text{GeV}\). This trend is observed across all fluctuation measures. This could possibly indicate a transition from baryon to hadron degrees of freedom or deconfinement, previously seen in other observables, now manifesting in fluctuation measures.

\section{Results}
The scaled variance, skewness, and kurtosis measured by the Beam Energy Scan II at RHIC are compared to two models that distinguish baryon charge fluctuations from two sources: stopping and pair creation. The stopped charge may be carried by either baryons or quarks. The nature of the charge carrier significantly affects fluctuations, especially at lower energies where stopped charge dominates particle production at mid-rapidity. At higher energies, pair creation becomes the primary source of fluctuations, rendering the nature of charge stopping less significant.

Since only hadrons, not quarks, are observed in the final state, my analysis employs a recombination model where hadrons are formed from quarks. This model conserves charge fluctuations and allows hadrons to move independently in rapidity space. Each hadron’s mobility results in fluctuations that equilibrate to values expected for a hadron gas. Due to the finite collision time, this equilibration is partial, and some quark-phase signals can still be observed. Notably, these signals are more pronounced in higher-order cumulants.

Comparing the model to data indicates that baryon stopping dominates at lower collision energies, with quark stopping becoming evident around $\sqrt{s_{NN}} \approx 20~\text{GeV}$. This shift explains the deviation in kurtosis from the global conservation baseline and coincides with the onset of deconfinement observed in other measurements.

\begin{acknowledgments}
The author is thankful to S. Pratt and M. Gorenstein, V.Koch, V.Vovchenko, O.Stashko for fruitful comments and discussions. 
Supported by the Department of Energy Office of Science through Grant No. DE-FG02-03ER41259. 
\end{acknowledgments}

\bibliography{main.bib}

\end{document}